\documentclass[twocolumn,showpacs,preprintnumbers,amsmath,amssymb,pra]{revtex4}
\usepackage{amsfonts}
\usepackage{mathrsfs}
\usepackage{graphicx}
\usepackage{dcolumn}
\usepackage{bm}

\begin{document}

\title{Entanglement spectrum: Identification of the transition
           from vortex-liquid to vortex-lattice state in a
           weakly interacting rotating Bose-Einstein condensate}

\author{Zhao Liu$^1$, Hong-Li Guo$^1$, Vlatko Vedral$^{2,3,4}$ and Heng Fan$^1$}
\affiliation{%
$^1$Institute of Physics, Chinese Academy of Sciences, Beijing
100190, China\\
$^2$Clarendon Laboratory, University of Oxford, Parks Road, Oxford OX1 3PU, United Kingdom \\
$^3$Centre for Quantum Technologies, National University of
Singapore, 3
Science Drive 2, Singapore 117543 \\
$^4$Department of Physics, National University of Singapore, 2
Science Drive 3, Singapore 117542
}%
\date{\today}

\begin{abstract}
We use entanglement to investigate the transition from vortex liquid
phase to vortex lattice phase in weakly interacting rotating
Bose-Einstein condensate (BEC). For the torus geometry, the ground
state entanglement spectrum is analyzed to distinguish these two
different phases. The low-lying part of ground state entanglement
spectrum, as well as the behavior of its lowest level change clearly
when the transition occurs. For the sphere geometry, the
entanglement gap in the conformal limit (CL) is also studied. We
also show that the decrease of entanglement between particles can be
regarded as a signal of the transition.
\end{abstract}

\maketitle


\section{Introduction}Recently trapped rotating ultracold
atoms have attracted considerable interest \cite{Bloch,ALF}. When
the rotating frequency is moderate, the formation of vortex lattice
is observed in experiments. However, with the increase of rotating
frequency, it is predicted that vortex lattice will melt and the
system will enter into a regime of strongly correlated many-body
physics, namely vortex liquid phase \cite{NJ,NNJ}. The ground states
in this liquid phase are the bosonic version of quantum Hall states,
leading to the invalidity of mean field theory
\cite{NK,HSK,SV,DD,Liu}. The parameter controlling this quantum
phase transition from vortex lattice phase to vortex liquid phase is
the average filling factor $\nu\equiv N/N_{V}$, where $N$ is the
number of bosons and $N_{V}$ is the number of vortices. Exact
diagonalization on the torus geometry provides an evidence that this
transition occurs at the critical filling factor $\nu_{c}\sim6$. For
$\nu>\nu_{c}$, the ground state is a vortex lattice while for
$\nu<\nu_{c}$, the ground state is a strongly correlated liquid
\cite{NNJ}.

The description of condensed matter phases using quantities such as
entanglement entropy and fidelity borrowed from quantum information
theory is another promising research field attracting great
attention \cite{Vidal}. It has been found that when probing
topologically ordered states, topological entanglement entropy,
which is the sub-leading term of the ground state entanglement
entropy, can provide some information that cannot be obtained by
conventional condensed matter methods \cite{Kit,Wen}. Quantum Hall
states of two-dimensional electrons in magnetic field and
two-dimensional bosons in rotating traps are very important
topologically ordered states, whose topological entanglement
entropies are calculated recently \cite{Haq,AGD}. Comparing with
generally used entanglement entropy which is only a single number,
the entanglement spectrum (ES) contains more information and is
found very recently to be useful in probing whether a quantum state
has topological order \cite{HLi}. In these two years, a series of
papers focus on the ES of fermionic fractional quantum Hall states
on the sphere geometry and torus geometry \cite{HLi,Haq2,AM}, the ES
of topological insulators and superconductors \cite{LF}, the ES of
spin model such as Heisenberg model and Kitaev model
\cite{DP,DPA,YH} and the complete definition of entanglement gap in
ES \cite{RT}. However, the ES of rotating BEC still lacks enough
study especially for large filling factors beyond quantum Hall
regime and we will study it in this work.

The ES is not only useful for systems of condensed matter physics.
It is also important in quantum information theory. By majorization
scheme of the ES, we can find deterministically whether a bipartite
quantum state can be transformed to another by local quantum
operations and classical communication while the entanglement
entropy is generally not sufficient \cite{Nielsen}.

In this work, we analyze the ground state ES of rotating BEC to
investigate the transition from vortex lattice phase to vortex
liquid phase from an entanglement perspective. Exact diagonalization
method is used to obtain the ground state on the torus and sphere
geometry. Then the ES is extracted from the ground state. Comparing
with Ref. \cite{NNJ}, where the excitation energy is used to
identify the phase transition, we just utilize ground state
properties, namely ES to indicate it. For the torus geometry, when
bosons interact through contact interaction, we find that with the
increase of filling factor, the low-lying part of ground state ES as
well as the behavior of its lowest level undergo a qualitative
change. The critical filling factor where this change happens is
just $\nu_{c}\sim6$. This phenomenon strongly indicates that a
transition of ground state indeed occurs at $\nu_{c}\sim6$. This is
consistent with the conclusion of Ref. \cite{NNJ}. When we change
the interaction to Coulomb potential, we find that the transition
occurs at a little larger $\nu_{c}$. For the sphere geometry and
contact interaction, the evolution of the entanglement gap in CL
with filling factors is studied. Finally, the entanglement between
two bosons and other $N-2$ bosons is also calculated. When
$\nu>\nu_{c}$, this entanglement decreases monotonically with $\nu$,
implying the transition from a strongly correlated regime to a mean
field regime.

\section{Model and method}Let us begin with a
brief introduction of the definition of entanglement spectrum.
Suppose we spatially divide a system into two parts $A$ and $B$,
then any pure state $|\Psi\rangle_{AB}$ of the system can be
expressed in Schmidt decomposition
$|\Psi\rangle_{AB}=\sum_{i}e^{-\xi_{i}/2}|\phi_{i}\rangle_{A}|\varphi_{i}\rangle_{B}$,
where $\sum_{i}e^{-\xi_{i}}=1$, $\xi_{i}\geq0$, and
$\{|\phi_{i}\rangle_{A}\}$ ($\{|\varphi_{i}\rangle_{B}$\}) is an
orthonormal basis of subsystem $A$ ($B$). The reduced density
operator of $A$ is $\rho_{A}=\exp(-\widetilde{H})$ with
$\widetilde{H}=\sum_{i}\xi_{i}|\phi_{i}\rangle_{A}\langle\phi_{i}|$.
Considering in statistical physics a density operator can be related
with Hamiltonian $\mathcal {H}$ through $\rho=e^{-\beta\mathcal
{H}}/\textrm{Tr}(e^{-\beta\mathcal {H}})$, We can regard
$\widetilde{H}$ as a pseudo-Hamiltonian and define $\{\xi_{i}\}$ as
entanglement spectrum.

In this paper, we first study a rotating $N$-boson system on the
torus with periods $a$ and $b$ in the $x$ and $y$ directions. Each
boson has a mass $m$. A consistent imposition of periodic boundary
conditions requires that $ab/\ell^{2}=2\pi N_{V}$, where
$\ell=\sqrt{\hbar/(2m\omega)}$ with $\omega$ the rotating frequency
and $N_{V}$ is the number of vortex. When the interaction is weak
enough, the lowest Landau level approximation is valid. The
normalized single-particle lowest Landau level (LLL) wavefunction is
\begin{eqnarray}
\psi_{j}=\Big(\frac{1}{a\ell\pi^{1/2}}\Big)^{1/2}\sum_{n=-\infty}^{+\infty}\exp\Big[\textrm{i}\Big(\frac{2\pi
j}{a}+\frac{nb}{\ell^{2}}\Big)x\nonumber\\
-\frac{1}{2\ell^{2}}\Big(y+nb+\frac{2\pi
j}{a}\ell^{2}\Big)^{2}\Big],\nonumber
\end{eqnarray}
where $j=0,1,2,...,N_{V}-1$. Because $\psi_{j}$ is centered along
the line $y=-\frac{2\pi j}{a}\ell^{2}$, the whole system can be
divided into $N_{V}$ localized orbits spatially. The interaction
Hamiltonian $V(\textbf{r}_{1},\textbf{r}_{2})$ after standard second
quantization procedure becomes
$V=\sum_{k_{1}k_{2}k_{3}k_{4}=0}^{N_{V}-1}V_{k_{1}k_{2}k_{3}k_{4}}
a_{k_{1}}^{\dagger}a_{k_{2}}^{\dagger}a_{k_{3}}a_{k_{4}}$, where
$a_{k}$ ($a_{k}^{\dagger}$) annihilates (creates) a boson in the
state $\psi_{k}$. On the torus, the matrix element
$V_{k_{1}k_{2}k_{3}k_{4}}$ takes the form as
\begin{eqnarray}
V_{k_{1}k_{2}k_{3}k_{4}}=\frac{\delta'_{k_{1}+k_{2},k_{3}+k_{4}}}{2ab}\sum_{s=-\infty}^{+\infty}\sum_{t=-\infty}^{+\infty}
\delta'_{k_{1}-k_{4},s}\widetilde{V}\Big(\frac{2\pi s}{a},\frac{2\pi
t}{b}\Big)\nonumber\\
\exp\Big\{-\frac{\ell^{2}}{2} \Big[\Big(\frac{2\pi s}{a}\Big)^{2}+
\Big(\frac{2\pi t}{b}\Big)^{2}\Big]
-2\pi\textrm{i}t\frac{k_{1}-k_{3}}{N_{V}}\Big\},\nonumber
\end{eqnarray}
where $\delta'$ is the periodic Kronecker delta function with period
$N_{V}$ and $\widetilde{V}(q_{x},q_{y})$ is the Fourier
transformation of interaction $V(\textbf{r})$ \cite{DY,EW}. For a
Fock basis $|n_{0},n_{1},...,n_{N_{V}-1}\rangle$ satisfying
$\sum_{j=0}^{N_{V}-1}n_{j}=N$, one can find that there is a
conserved quantity $K=\sum_{j=0}^{N_{V}-1}jn_{j}$ (mod $N_{V}$),
which can be called total momentum. The Lanczos algorithm is used to
diagonalize the interaction matrix $V$ to calculate the ground state
energy and ground state wavefunction in each sector of $K$ and then
the lowest energy $E_{0}$ and its eigenstate $|\Phi_{0}\rangle$ are
selected. To extract the ES from $|\Phi_{0}\rangle$, we bipartition
the system into blocks $A$ and $B$ which consist of $l_{A}$
consecutive orbits and the remaining $N_{V}-l_{A}$ orbits
respectively. In this work, we focus on the case where
$l_{A}=N_{V}/2$, namely $A$ contains the orbits from 0 to
$N_{V}/2-1$ and $B$ contains the orbits from $N_{V}/2$ to $N_{V}-1$.
ES is labeled by the particle number $N_{A}=\sum_{j\in A}n_{j}$ and
the total momentum $K_{A}=\sum_{j\in A}jn_{j}$ (mod $N_{V}$) in
block $A$.

Similarly, we can consider a $N$-bosons system on a sphere surface. Then under the
LLL approximation, the contact interaction Hamiltonian can be written as $V=\sum_{k_{1}k_{2}k_{3}k_{4}=0}^{2S}V_{k_{1}k_{2}k_{3}k_{4}}
a_{k_{1}}^{\dagger}a_{k_{2}}^{\dagger}a_{k_{3}}a_{k_{4}}$ with
\begin{eqnarray}
V_{k_{1}k_{2}k_{3}k_{4}}=\delta_{k_{1}+k_{2},k_{3}+k_{4}}
\frac{1}{2S}\frac{(2S+1)^{2}}{4S+1}\frac{[\prod_{i=1}^{4}\textrm{C}_{k_{i}}^{2S}]^{1/2}}{\textrm{C}_{k_{1}+k_{2}}^{4S}},\nonumber
\end{eqnarray}
where $\textrm{C}_{m}^{n}=\frac{n!}{(n-m)!m!}$ and $N_{V}=2S$.
$a_{k}$ ($a_{k}^{\dagger}$) annihilates (creates) a boson with
$z$-component angular momentum $k$. ES is labeled by the particle
number $N_{A}=\sum_{j\in A}n_{j}$ and the total $z$-component
angular momentum $L^{A}_{z}=\sum_{j\in A}jn_{j}$ in block $A$.

\section{Entanglement spectrum}
Before we study the transition from vortex liquid to vortex lattice,
we want to show some ES data of typical bosonic quantum Hall states.
We use two-body contact interaction
$\sum_{i<j}\delta(\textbf{r}_{i}-\textbf{r}_{j})$ and three-body
contact interaction
$\sum_{i<j<k}\delta(\textbf{r}_{i}-\textbf{r}_{j})\delta(\textbf{r}_{j}-\textbf{r}_{k})$
to obtain the exact Laughlin state at $\nu=1/2$ and Pfaffian state
at $\nu=1$ respectively. Throughout our calculation we take
$\hbar/(m\omega)=1$, leading to $\ell^{2}=1/2$. When plotting the ES
data, we choose the sector of $N_{A}$ where the lowest ES level is
found. Then in this $N_{A}$ sector, the $K_{A}$ sector where the
lowest ES level locates is put in the center of the spectrum.
$\Delta K_{A}$ is defined as the shift of momentum from the center
of the spectrum. For example, for $N_{V}=6$, the possible values of
$K_{A}$ are 0,1,2,3,4 and 5. If we find the lowest ES level locates
in $K_{A}=2$ sector, then we plot ES data according to the order
$K_{A}=5,0,1,2,3,4,5$ from left to right to put $K_{A}=2$ sector in
the center of the spectrum, corresponding to $\Delta
K_{A}=-3,-2,-1,0,1,2,3$.

\begin{figure}
\includegraphics[height=4cm,width=0.8\linewidth]{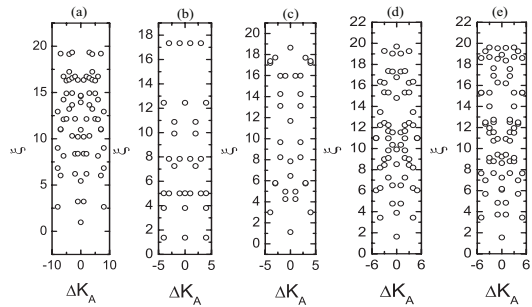}
\caption{\label{v1} The ES of typical quantum Hall states (the
aspect ratio is 1). The Laughlin state is two-fold degenerate on the
torus: one is in sector $K=0$ and the other is in sector
$K=N_{V}/2$. The Pfaffian state is three-fold degenerate on the
torus: two are in sector $K=0$ and one is in sector $K=N_{V}/2$. (a)
The ES of Laughlin state at $\nu=1/2$ for $N=8,N_{V}=16$ in sector
$K=0$. (b) The ES of Pfaffian state at $\nu=1$ for $N=8,N_{V}=8$ in
sector $K=0$. (c) The ES of Pfaffian state at $\nu=1$ for
$N=8,N_{V}=8$ in sector $K=4$. (d) The ES of Pfaffian state at
$\nu=1$ for $N=10,N_{V}=10$ in sector $K=0$. (e) The ES of Pfaffian
state at $\nu=1$ for $N=10,N_{V}=10$ in sector $K=5$.}
\end{figure}

The shapes of the ES of both Laughlin and Pfaffian states can be
explained by considering their thin torus limit ($a\rightarrow0$).
We give some examples of such an analysis below. For the Laughlin
state with $N=2, N_V=4$ in $K=0$ sector, the thin torus state is
$|0101\rangle$, so the reduced density operator of part $A$ is
$\rho_A=|01\rangle\langle01|$ leading to only one level in the
bottom of ES in $N_A=1$ sector. When we enlarge $a$, more ES levels
come down and a V-shape appears [Fig.\ref{v1}(a)]. For fermion
Laughlin state, this V-shape can be explained by two chiral edge
theory for a properly small $a$ depending on the system size
\cite{AM}. For the Pfaffian state with $N=N_V=4$ in $K=0$ sector,
the thin torus states are $(|0202\rangle\pm|2020\rangle)/\sqrt{2}$,
so the reduced density operator of part $A$ is
$\rho_A=(|02\rangle\langle02|+|20\rangle\langle20|)/2$, leading to
two equal ES levels in $N_A=2$ sector with $K_A=0$ and 2
respectively. So the ES has a flat bottom with three degenerate
levels in $\Delta K_A=0, \pm2$. When we enlarge $a$, more levels
come down but the flat bottom with three degenerate levels is still
there [Fig.\ref{v1}(b)]. However, this flat bottom can disappear if
we choose another system size. For the Pfaffian state with $N=N_V=6$
in $K=0$ sector, the thin torus states are
$(|020202\rangle\pm|202020\rangle)/\sqrt{2}$, so the reduced density
operator of part $A$ is
$\rho_A=(|020\rangle\langle020|+|202\rangle\langle202|)/2$, leading
to two equal ES levels again. But they are in different $N_A$
sectors ($N_A=2$ and $N_A=4$ respectively). So, when plotting the ES
in one $N_A$ sector, the flat bottom disappears and V-shape revives
[Fig.\ref{v1}(d)]. At last, for the Pfaffian state with $N=N_V=4$ in
$K=2$ sector, the thin torus state is $|1111\rangle$. So it is easy
to see there is only one ES level in the bottom in $N_A=2$ sector,
meaning a V-shape will appear when $a$ increases
[Fig.\ref{v1}(c,e)].

In Ref. \cite{NNJ}, a trial wavefunction for vortex liquid phase at
$\nu=k/2$ with $k$ an integer is proposed:
\begin{eqnarray}
\Psi^{k}=\mathcal {S}\Big[\prod_{i<j\in
A_{1}}^{N/k}(z_{i}-z_{j})^{2}...\prod_{l<m\in
A_{k}}^{N/k}(z_{l}-z_{m})^{2}\Big],\nonumber
\end{eqnarray}
where we express it in disk geometry and the Gaussian exponential
factor is omitted. In every set $A_i$, there are $N/k$ bosons and
$\mathcal {S}$ means the symmetrization over all possible partition
of the $N$ bosons into $k$ sets. When $k=1(2)$, this state is
Laughlin (Pfaffian) state. We expect that the ES of incompressible
quantum Hall states at other filling factors on the torus have
similar shapes with Laughlin and Pfaffian states. Therefore if we
see a V-shape or a flat bottom with three degenerate levels in the
ES of one state, we can conjecture that this state has a close
relationship with quantum Hall state.

\begin{figure*}
\includegraphics[height=8cm,width=0.6\linewidth]{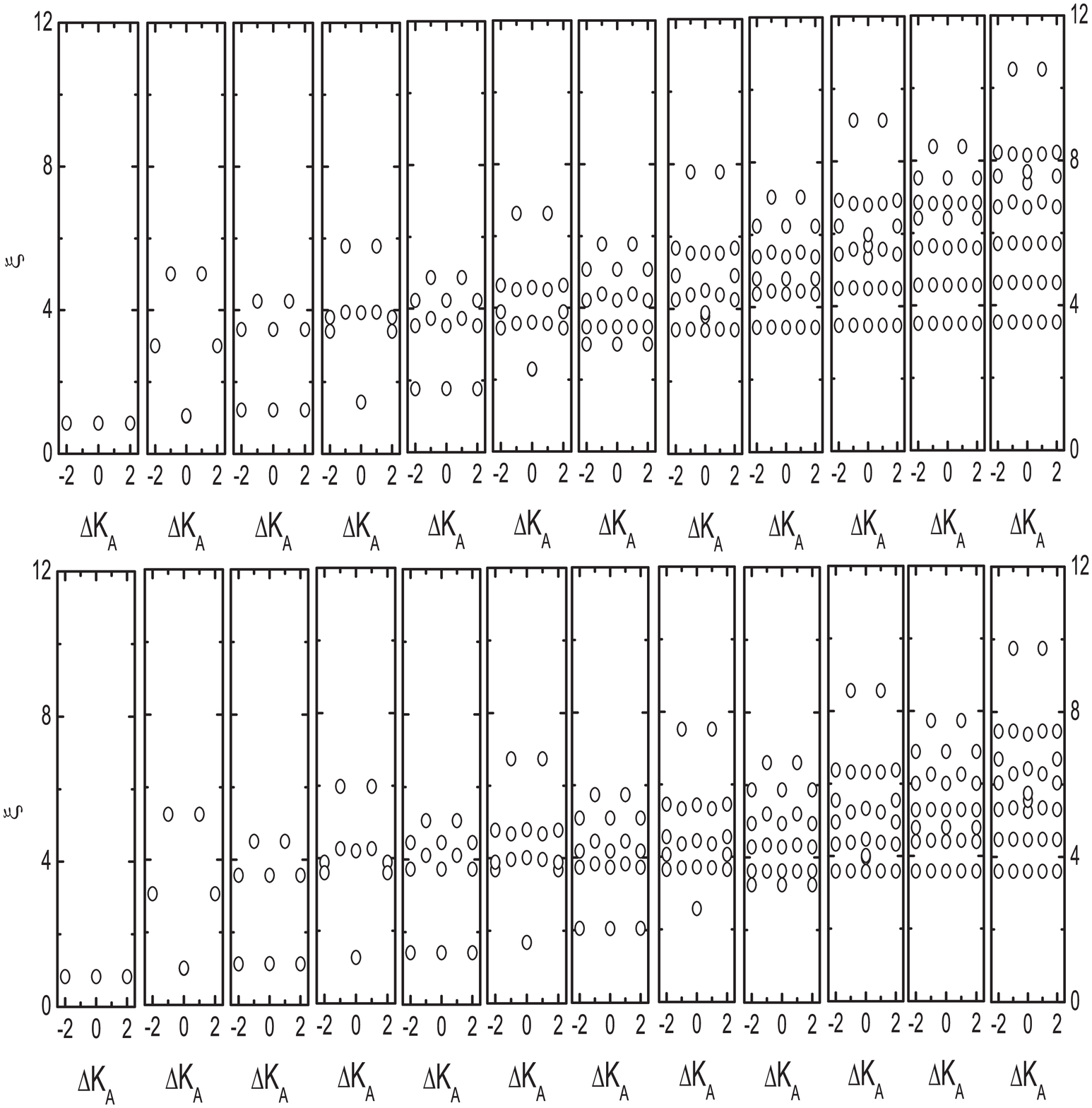}
\caption{\label{int4} The ground state ES for $N_{V}=4$ at integer
filling factor $\nu=1,2,...,11,12$ from left to right. The subsystem
$A$ consists of orbits 0 and 1 and the subsystem $B$ consists of
orbits 2 and 3. The $N_{A}$ sector is chosen as the one where the
lowest ES level (corresponding to the largest eigenvalue of
$\rho_{A}$) locates. The aspect ratio is chosen as $\sqrt{3}/2$. The
interaction can be either contact interaction (the first row) or
Coulomb interaction (the second row). }
\end{figure*}

\begin{figure*}
\includegraphics[height=8cm,width=0.6\linewidth]{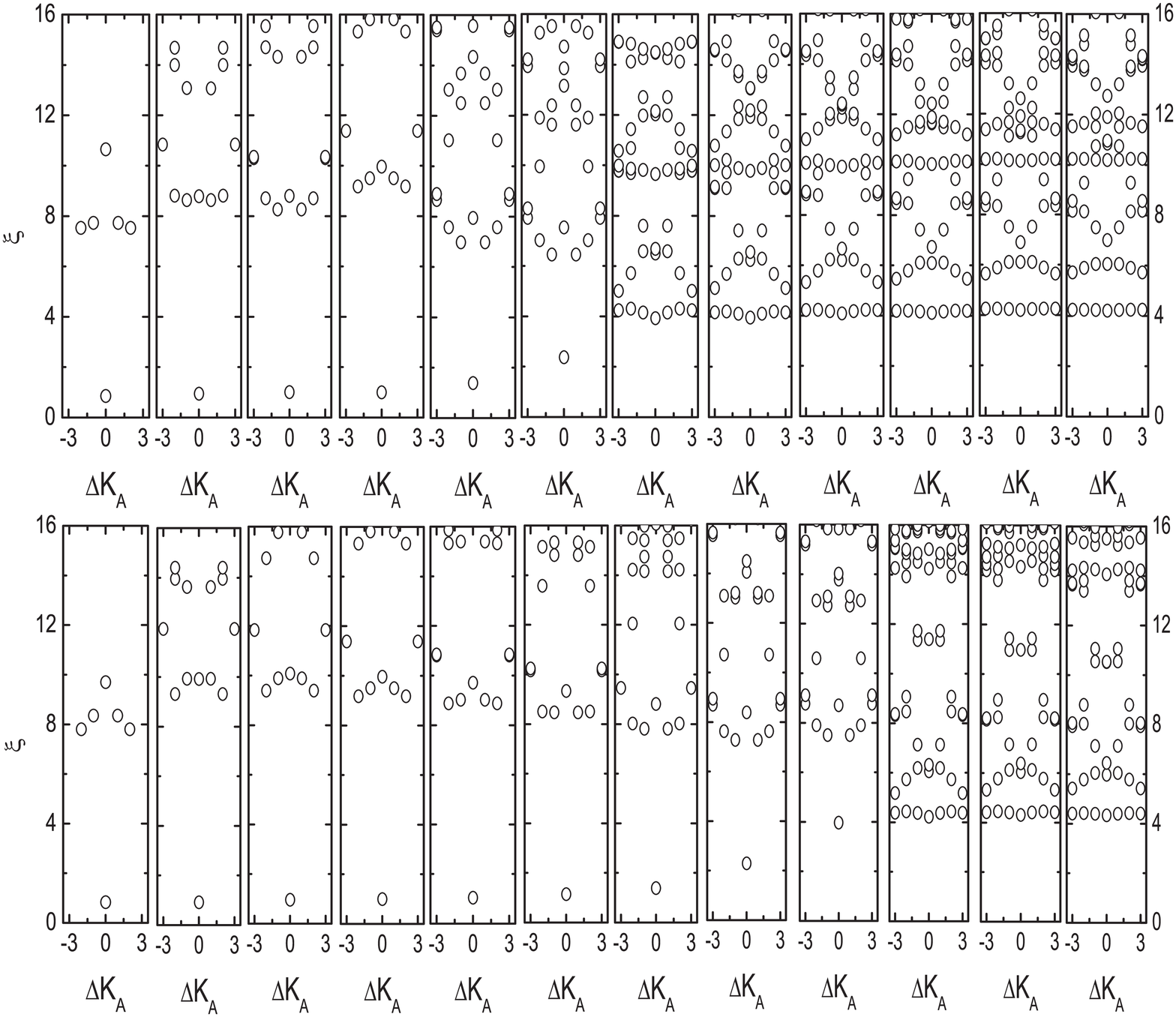}
\caption{\label{int6} The ground state ES for $N_{V}=6$ at integer
filling factor $\nu=1,2,...,11,12$ from left to right. The subsystem
$A$ consists of orbits 0, 1 and 2 and the subsystem $B$ consists of
orbits 3, 4 and 5. The $N_{A}$ sector is chosen as the one where the
lowest ES level (corresponding to the largest eigenvalue of
$\rho_{A}$) locates. The aspect ratio is chosen as $1/\sqrt{3}$. The
interaction can be either contact interaction (the first row) or
Coulomb interaction (the second row).}
\end{figure*}

Now we consider a system on the torus with the aspect ratio $a/b$
commensurate with a triangular lattice, which can help us to
stabilize the vortex lattice. We suppose the interactions are
contact interaction $V(\textbf{r})=\delta(\textbf{r})$ and Coulomb
interaction $V(\textbf{r})=1/|\textbf{r}|$, for which
$\widetilde{V}(q_{x},q_{y})\propto1$ and
$\widetilde{V}(q_{x},q_{y})\propto(q_{x}^{2}+q_{y}^{2})^{-1/2}$
respectively. Because we will consider the ground state at large
filling factors, the number of vortex cannot be too large. First we
study the cases at integer filling factors. When $\nu$ is a small
integer (e.g., $\nu\lesssim10$ for $N_{V}=4$), the ground state is
unique and for our choice of aspect ratio it is in $K=0$ sector.
When $\nu$ is a large integer (deeply in vortex lattice phase, e.g.,
$\nu\gtrsim10$ for $N_{V}=4$), the ground state energies in
different $K$ sectors become nearly the same. After checking we find
all the ES of these nearly-degenerate ground states have
qualitatively the same structure, so we still choose the ground
state in $K=0$ sector to extract the ES. In Figs.\ref{int4} and
\ref{int6}, we plot the ground state ES for $N_{V}=4$ and $N_{V}=6$
at integer filling factors $\nu=1,2,...,11,12$ respectively. For
$N_{V}=6$, let us first consider contact interaction. One can see
that the qualitative behavior of the low-lying part of ES clearly
changes after $\nu=6$, distinguishing the vortex lattice phase and
vortex liquid phase. For $\nu\leq6$, the ground state ES has a
structure of V-shape, while for $\nu\geq7$, this V-shape structure
vanishes and the ground levels of ES in every $\Delta K_{A}$ sector
have a tendency to become degenerate, implying a transition from
quantum Hall region. If we change the interaction to Coulomb
interaction, the structure of the ground state ES changes after a
little larger filling factor $\nu=9$. Similar phenomena occur for
$N_{V}=4$. The qualitative behavior of the low-lying part of the ES
clearly changes after $\nu=7$ and $\nu=9$ for contact interaction
and Coulomb interaction respectively. Before the transition, the
shape of the ES is either a V-shape (at even filling factors) or a
flat bottom with three degenerate levels (at odd filling factors).
After the transition, the ground levels of ES in every $\Delta
K_{A}$ sector have a tendency to become degenerate.

When the filling factor is a fraction $p/q$ where $q\neq1$ and $p$
and $q$ are coprime, the ground state is at least $q$-fold
degenerate. For our choice of aspect ratio, there is always one of
these degenerate ground states in $K=0$ sector. After calculating
the ES of all degenerate ground states (e.g., for $q=2$), we find
that just like in the case of integer filling factors, with the
increase of $\nu$, the ground levels of the ES in every $\Delta
K_{A}$ sector have a tendency to become degenerate.


\begin{figure}
\includegraphics[height=3.5cm,width=\linewidth]{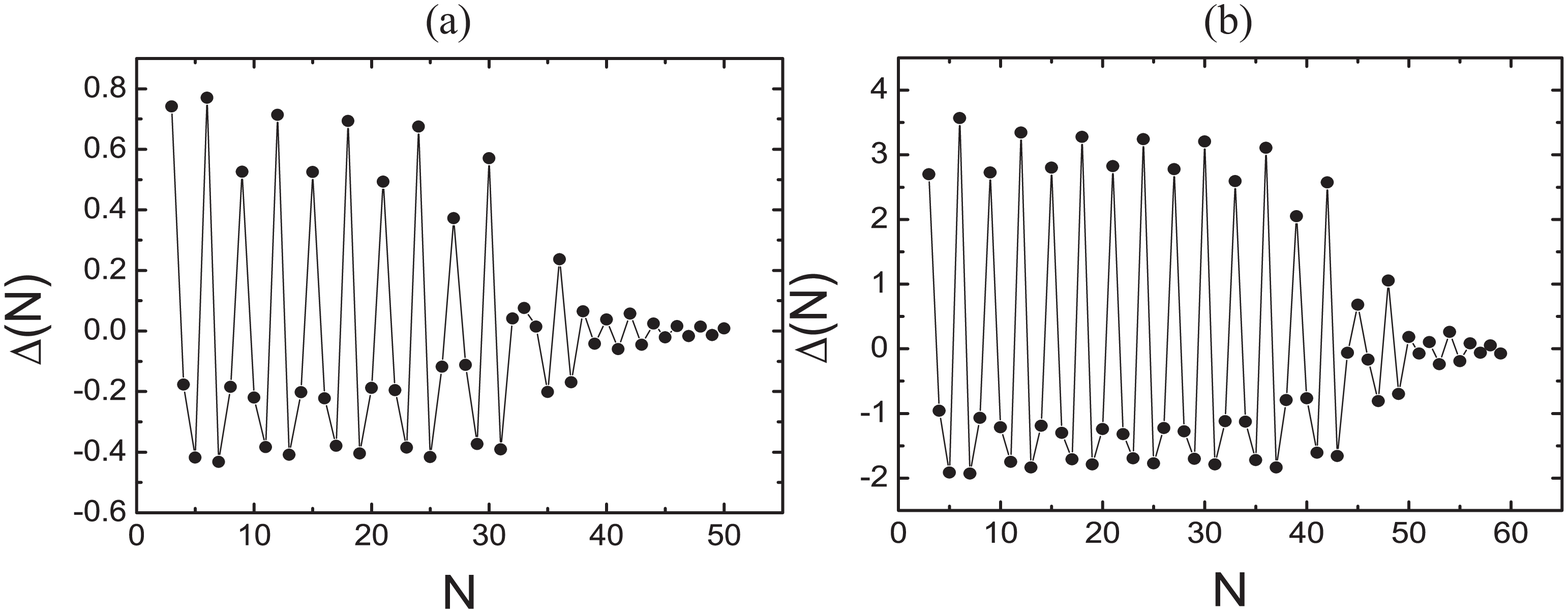}
\caption{\label{gap} The ground energy gap $\Delta(N)$ of the
Hamiltonian as a function of $N$ for (a) contact interaction and (b)
Coulomb interaction. $\Delta(N)$ reflects the discontinuity of
chemical potential and is defined as
$\Delta(N)=N[E(N+1)/(N+1)+E(N-1)/(N-1)-2E(N)/N]$, where $E(N)$ is
the ground energy of the system consisting of $N$ particles. The
vanish of $\Delta(N)$ indicates the transition from vortex liquid to
compressible vortex lattice. Here $N_{V}=6$ and the aspect ratio is
$1/\sqrt{3}$. The unit of the energy is $\hbar\omega$.}
\end{figure}

In Ref. \cite{NNJ}, a ground energy gap of the Hamiltonian is
defined which can indicate the transition from vortex liquid to
vortex lattice (Fig.\ref{gap}) [Fig.\ref{gap}(a) can also be found
in Ref. \cite{NNJ}]. One can see for $N_V=6$ the ground energy gap
vanishes after $\nu=6$ and $\nu=8$ for contact and Coulomb
interaction respectively, which is consistent with the conclusion
obtained by ground state ES.

\begin{figure}
\includegraphics[height=3.5cm,width=\linewidth]{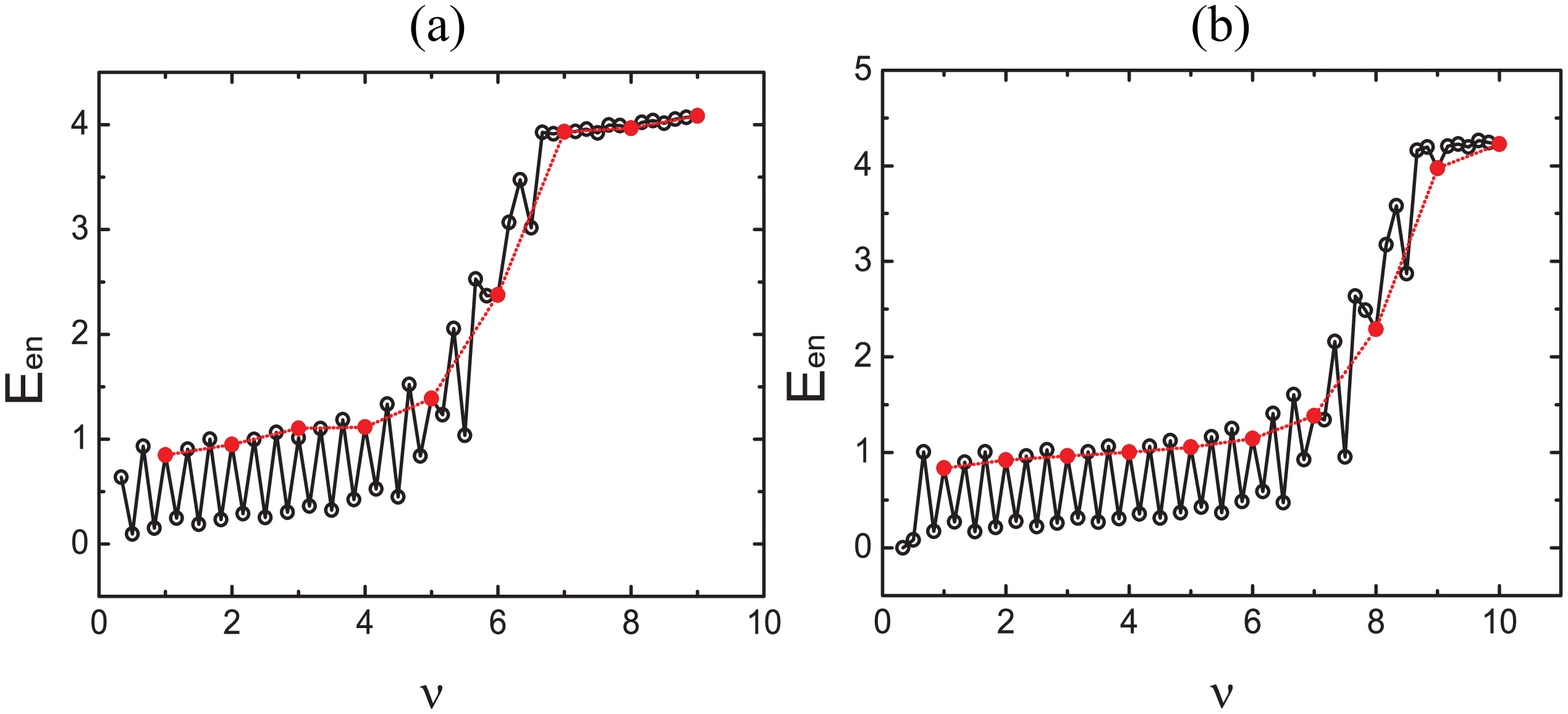}
\caption{\label{eend} (color online) The lowest level
$E_{\textrm{en}}$ of the ground state ES as a function of $\nu$ for
(a) contact interaction and (b) Coulomb interaction. When the ground
state is degenerate (or nearly-degenerate), we choose the ground
state in $K=0$ sector to extract the ES data. After checking, we
find this choice will not influence our results qualitatively. The
integer filling factors are emphasized by red dots. Here $N_{V}=6$
and the aspect ratio is $1/\sqrt{3}$.}
\end{figure}

Considering the critical point of a quantum phase transition can be
indicated by the singular behavior of the ground energy (or its
derivative) of the physical Hamiltonian and the ES corresponds to
the spectrum of a pseudo-Hamiltonian $\widetilde{H}$, we hope to see
the ground energy of ES can indicate the transition from vortex
liquid to vortex lattice. We investigate the evolution of the lowest
level $E_{\textrm{en}}$ of ground state ES with filling factor
$\nu$. Here we take a system containing six vortices as an example.
For contact interaction, from Fig.\ref{eend}(a) we can see when
$\nu\leq5$, $E_{\textrm{en}}$ increases slowly and oscillates
obviously with $\nu$. When $\nu\geq7$, $E_{\textrm{en}}$ also
increases very slowly with $\nu$ but stops clear oscillation. A
transition regime can be found at $5<\nu<7$, in which
$E_{\textrm{en}}$ goes up steeply with the increase of $\nu$,
showing a singular behavior like that in first-order quantum phase
transition. Similar phenomena happen for Coulomb interaction
(Fig.\ref{eend}(b)).

We wonder wether the transition from a bosonic quantum Hall state to
a compressible state can also be reflected in other geometry. In
Ref. \cite{RT}, the ES in the conformal limit (CL) on the sphere
geometry is defined. The spirit of CL is to unnormalize the ground
state obtained from the diagonalization of the Hamiltonian by
expressing the ground state in a new set of basis states which do
not contain the normalization factor of single-particle LLL wave
function, because this factor contains the information about the
extent of the Landau orbits in space and depends on the magnetic
length of the problem. On the sphere, the normalization factor of
$j$-th single-particle LLL is $\mathcal
{N}_{j}\propto\sqrt{(2S)!/[j!(2S-j)!]},j=0,1,...,2S$. If the ground
state is
$|\Psi\rangle=\sum_{n_{0},n_{1},...,n_{2S}}\psi_{n_{0},n_{1},...,n_{2S}}|n_{0},n_{1},...,n_{2S}\rangle$,
then after unnormalization, the ground state is expressed as
$|\widetilde{\Psi}\rangle\propto\sum_{n_{0},n_{1},...,n_{2S}}(\psi_{n_{0},n_{1},...,n_{2S}}\prod_{j=0}^{2S}\mathcal
{N}_{j}^{n_{j}})\widetilde{|n_{0},n_{1},...,n_{2S}\rangle}$, where
$\widetilde{|n_{0},n_{1},...,n_{2S}\rangle}$ is the new basis state
that does not contain $\mathcal {N}_{j}$. The ES in CL is extracted
from $|\widetilde{\Psi}\rangle$. For an incompressible quantum Hall
state, the ES in CL can be divided into two parts, one of which is
the low-lying conformal field part and the other is the generic part
caused by realistic interaction. There is an entanglement gap
between two parts \cite{HLi,RT}. If a state undergoes a transition
from a quantum Hall state to a compressible state, this entanglement
gap closes \cite{Haq2,RT}.

\begin{figure*}
\includegraphics[height=4cm,width=\linewidth]{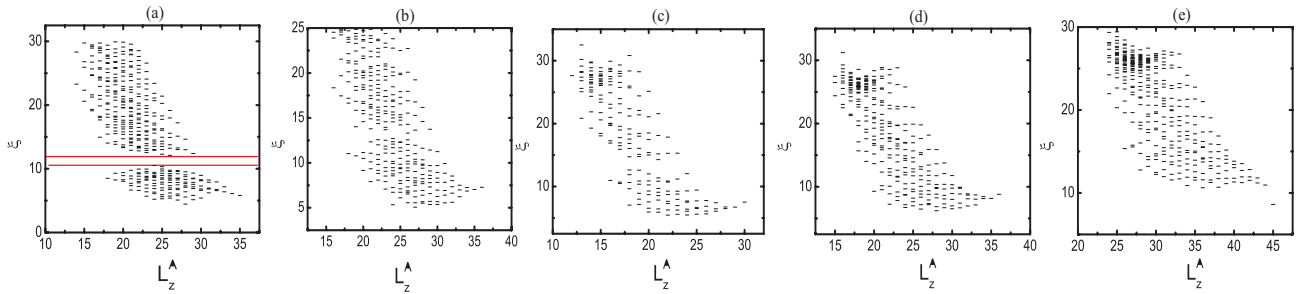}
\caption{\label{sphere} (color online) The ground state ES in CL on
the sphere geometry for contact interaction with filling factors
$\nu=1,3/2,2,5/2$ and 3 from left to right. Here $N=(N_{V}+2)\nu$.
Subsystem $A$ consists of orbits from 0 to $N_{V}/2-1$ and subsystem
$B$ consists of orbits from $N_{V}/2$ to $N_{V}$. Unlike what we do
on the torus geometry, we choose the sector $N_{A}=N/2$ for even $N$
[$N_A=(N-1)/2$ for odd $N$] to plot ES when considering sphere
geometry. (a) $\nu=1$, $N=14$, $N_{V}=12$. (b) $\nu=3/2$, $N=18$,
$N_{V}=10$. (c) $\nu=2$, $N=20$, $N_{V}=8$. (d) $\nu=5/2$, $N=25$,
$N_{V}=8$. (e) $\nu=3$, $N=30$, $N_{V}=8$. The entanglement gap is
clear at $\nu=1$ (indicated by the red lines), but hard to identify
for larger $\nu$.}
\end{figure*}

Here we study the evolution of ground state ES in CL with the
increase of filling factor on the sphere geometry
(Fig.\ref{sphere}). When the filling factor is small ($\nu=1$),
there is a clear entanglement gap separating the low-lying conformal
field theory part from the generic part. However, when the filling
factor becomes large ($\nu=3/2,2,5/2,3$), the gap is unclear and
hard to identify. Therefore there are not many hints for us to say
wether the ground state is an incompressible quantum Hall state or
not for $\nu\geq3/2$ in thermodynamic limit. Our conclusion obtained
from the entanglement gap is consistent with the one obtained from
the energy gap of the Hamiltonian \cite{NRTH}, where the authors
found for $\nu\geq3/2$ the energy gap is very small showing no clear
signs of convergence towards thermodynamic limit and the
two-particle correlation function shows very strong oscillations and
no hint of incompressibility. The possible reason is that the
quantum Hall states at larger filling factors have very large
correlation lengths which exceed our largest finite-size system.
Therefore, the ES, like the spectrum of the Hamiltonian, can detect
the quantum Hall states at small filling factors on the sphere.
However, probably it is helpful to observe a possible transition
from an incompressible quantum Hall state to a compressible state
only in much larger systems.

\section{Particle entanglement}
\begin{figure}
\includegraphics[height=4cm,width=0.7\linewidth]{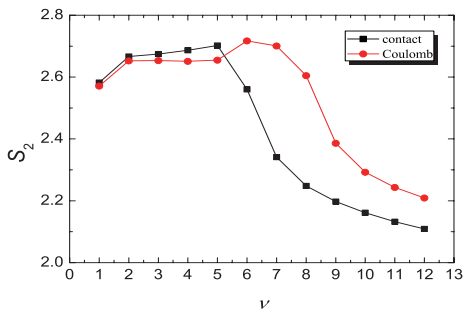}
\caption{\label{ent} (color online) The ground state particle
entanglement $S_{2}$ on the torus for both contact interaction
(black cubic) and Coulomb interaction (red circle). Here $N_{V}=6$
and the aspect ratio is $1/\sqrt{3}$.}
\end{figure}

We know that in
vortex lattice phase, the Gross-Pitaevskii mean field theory is
essentially correct. The many-body ground wavefunction in mean field
theory is just a product of wavefunctions of every particle, meaning
that in mean field theory there is no ground state entanglement
between particles. But in vortex liquid phase, the ground state is
strongly correlated. So it's natural to ask whether particle
entanglement can be an indicator of the transition from vortex
liquid to vortex lattice. Here we consider the ground state particle
entanglement between two bosons and other $N-2$ bosons, defined as
$S_{2}\equiv-\textrm{Tr}(\rho_{2}\ln\rho_{2})$. $\rho_{2}$ is the
reduced density operator of two bosons, whose matrix element is
$(\rho_{2})_{ij,kl}=\frac{1}{N(N-1)}\langle
a_{k}^{\dag}a_{l}^{\dag}a_{j}a_{i}\rangle_{\textrm{GS}}$, where
$\langle\bullet\rangle_{\textrm{GS}}$ is ground state average.
Considering the case $N_{V}=6$ on the torus, one can see that
$S_{2}$ decreases obviously after $\nu=6$ for contact interaction
and after $\nu=8$ for Coulomb interaction respectively
(Fig.\ref{ent}), giving a signal of the transition from vortex
liquid to vortex lattice, though not very exactly.

\section{Summary}In summary, we investigate the
quantum phase transition from vortex liquid phase to vortex lattice
phase in rotating BEC from an entanglement perspective. The ground
state entanglement spectrum is a useful tool in detecting this
transition. On the torus geometry its low-lying part has different
structures in vortex liquid phase and vortex lattice phase. Its
lowest level also behaves differently in these two phases. Therefore
ground state ES distinguishes these two phases very well. Our
results are consistent with the conclusion that for bosons
interacting through contact potential, a quantum phase transition
happens at $\nu_{c}\sim6$. When considering Coulomb interaction,
$\nu_{c}$ will become a little larger. On the sphere geometry, at
$\nu\geq3/2$, the entanglement gap of ES in CL is unclear, showing a
need to investigate much larger systems if one wants to observe the
melt of vortex lattice. Finally, we show that particle entanglement
can also be a signal to indicate the transition.

In this work, we choose the aspect ratio as the one commensurate
with triangular lattice. If we change the aspect ratio and the
interaction, we can study the ES of other possible ground states,
such as smectic state and vortex lattice state with other
symmetry. Moreover, it's also interesting to investigate
the ES of the simplest non-Abelian
bosonic Moore-Read state at $\nu=1$. In a very recent paper
\cite{AS}, a new kind of ES is defined where spatial bipartition of
the system is replaced by particle bipartition. We hope this new ES
can also indicate the transition from vortex liquid to vortex
lattice.

Current experiments are deep in the regime of vortex lattice, so it is still a
big challenge to create strongly-correlated states in vortex liquid phase.  Formation of these
new states requires faster rotation and smaller particle number than most current experiments.
However, some method to solve this problem has been proposed \cite{Fet}. We hope the transition
from vortex lattice to vortex liquid may be detected experimentally in the future.

\begin{acknowledgments}
We thank M. Haque, E. J. Bergholtz and N.
R. Cooper a lot for their very helpful discussions and comments.
This work is supported by NSFC grant (10974247) and ``973'' program
(2010CB922904).
\end{acknowledgments}

\newpage

\begin{thebibliography}{99}

\bibitem{Bloch}I. Bloch, J. Dalibard, and W. Zwerger, Rev. Mod.
Phys. {\bf 80}, 885 (2008).

\bibitem{ALF}A. L. Fetter, Rev. Mod. Phys. {\bf 81}, 647 (2009).

\bibitem{NJ}N. K. Wilkin and J. M. F. Gunn, Phys. Rev. Lett. {\bf 84}, 6
(2000).

\bibitem{NNJ}N. R. Cooper, N. K. Wilkin, and J. M. F. Gunn, Phys. Rev. Lett.
{\bf 87}, 120405 (2001).

\bibitem{NK}N. R. Cooper and E. H. Rezayi, Phys. Rev. A {\bf 75}, 013627 (2007).

\bibitem{HSK}Hidetsugu Seki and Kazusumi Ino, Phys. Rev. A {\bf 77}, 063602 (2008).

\bibitem{SV}S. Viefers, J. Phys.: Condens. Matter {\bf 20}, 123202
(2008).

\bibitem{DD}D. Dagnino, N. Barberan, M. Lewenstein, and J. Dalibard, Nat. Phys. {\bf 5}, 431 (2009).

\bibitem{Liu}Z. Liu, H. Guo, S. Chen and H. Fan, Phys. Rev. A {\bf
80}, 063606 (2009).

\bibitem{Vidal}L. Amico, R. Fazio, A. Osterloh, and V. Vedral, Rev.
Mod. Phys. {\bf 80}, 517 (2008).

\bibitem{Kit}A. Kitaev and J. Preskill, Phys. Rev. Lett. {\bf 96}, 110404 (2006).

\bibitem{Wen}M. Levin and X. G. Wen, Phys. Rev. Lett. {\bf 96}, 110405 (2006).

\bibitem{Haq}M. Haque, O. Zozulya, and K. Schoutens, Phys. Rev. Lett. {\bf 98}, 060401 (2007).

\bibitem{AGD}A. G. Morris, D. L. Feder, Phys. Rev. A {\bf 79},
013619 (2009).



\bibitem{HLi}H. Li and F. D. M. Haldane, Phys. Rev. Lett. {\bf 101}, 010504 (2008).

\bibitem{Haq2}O. Zozulya, M. Haque, and N. Regnault, Phys. Rev. B {\bf 79}, 045409 (2009).

\bibitem{AM}A. M. Lauchli, E. J. Bergholtz, J. Suorsa and M. Haque, Phys. Rev. Lett. {\bf 104}, 156404 (2010).

\bibitem{LF}L. Fidkowski, Phys. Rev. Lett. {\bf 104}, 130502 (2010).

\bibitem{DP}D. Poilblanc, Phys. Rev. Lett. {\bf 105}, 077202 (2010).

\bibitem{DPA}R. Thomale, D. P. Arovas and B. A. Bernevig, Phys. Rev. Lett. {\bf 105}, 116805 (2010).

\bibitem{YH}H. Yao and X.-L. Qi, Phys. Rev. Lett. {\bf 105}, 080501 (2010).

\bibitem{RT}R. Thomale, A. Sterdyniak, N. Regnault, and B. A. Bernevig, Phys. Rev. Lett. {\bf 104}, 180502 (2010).

\bibitem{Nielsen}M. A. Nielsen, Phys. Rev. Lett. {\bf 83}, 436 (1999).

\bibitem{DY}D. Yoshioka, B. I. Halperin and P. A. Lee, Phys. Rev.
Lett. {\bf 50}, 1219 (1983).

\bibitem{EW}E. Wikberg, E. J. Bergholtz and A. Karlhede, J. Stat. Mech.: Theory Exp. ({\bf 2009}), P07038.

\bibitem{NRTH}N. Regnault and Th. Jolicoeur, Phys. Rev. Lett. {\bf 91}, 030402 (2003).


\bibitem{AS}A. Sterdyniak, N. Regnault and B. A. Bernevig, arXiv: 1006. 5435.

\bibitem{Fet}A. L. Fetter, Rev. Mod. Phys. {\bf 81}, 647 (2009).

\end{thebibliography}
\end{document}